\documentclass[aps,pra,twocolumn,superscriptaddress]{revtex4-1}

\usepackage{amsmath}
\usepackage{amssymb}
\usepackage{epsfig}
\usepackage{color}
\usepackage{soul}
\usepackage{epstopdf}
\usepackage{subfigure}

\graphicspath{{.}{./figs/}}

\newcommand{\braket}[2]{\langle{#1}|{#2}\rangle}
\newcommand{\ketbra}[2]{|{#1}\rangle\langle{#2}|}
\newcommand* {\bra}[1]{\ensuremath{\langle {#1} |}}
\newcommand* {\ket}[1]{\ensuremath{| {#1} \rangle}}

\begin{document}

\title{
Measurement-induced chaos and quantum state discrimination in an iterated Tavis-Cummings scheme 
}

\author{Juan Mauricio Torres} %\email{mauricio.torres@physik.tu-darmstadt.de}
\affiliation{Institut f\"{u}r Angewandte Physik, Technische Universit\"{a}t Darmstadt, D-64289, Germany}
\affiliation{Instituto de F\'isica, Benem\'erita Universidad Aut\'onoma de Puebla, Apdo. Postal J-48, Puebla, Pue. 72570, M\'exico}
\author{J\'ozsef Zsolt Bern\'ad} 
\affiliation{Institut f\"{u}r Angewandte Physik, Technische Universit\"{a}t Darmstadt, D-64289, Germany}
\author{Gernot Alber} 
\affiliation{Institut f\"{u}r Angewandte Physik, Technische Universit\"{a}t Darmstadt, D-64289, Germany}
\author{Orsolya K\' alm\' an} 
\affiliation{
Institute for Solid State Physics and Optics, Wigner Research Centre, Hungarian Academy of Sciences, 
P.O. Box 49, H-1525
Budapest, Hungary}
\author{Tam\'as Kiss}
\affiliation{
Institute for Solid State Physics and Optics, Wigner Research Centre, Hungarian Academy of Sciences, 
P.O. Box 49, H-1525
Budapest, Hungary}
\date{\today}
\begin{abstract}
	A cavity quantum electrodynamical scenario is proposed for implementing a
	Schr\"odinger microscope capable of amplifying differences between non
	orthogonal atomic quantum states. The scheme involves an ensemble of
	identically prepared two-level atoms interacting pairwise with a single mode
	of the radiation field as described by the Tavis-Cummings model. By repeated
	measurements of the cavity field and of one atom within each pair a
	measurement-induced nonlinear quantum transformation of the relevant atomic
	states can be realized. The intricate dynamical properties of this nonlinear
	quantum transformation, which exhibits measurement-induced chaos, allows
	approximate orthogonalization of atomic states by purification after a few
	iterations of the protocol, and thus the application of the scheme for
	quantum state discrimination.
\end{abstract}
\maketitle
\section{Introduction}
Consistent with the no-cloning theorem nonorthogonal quantum states cannot
be distinguished perfectly. However, for purposes of quantum communication,
for example, it is necessary to be able to distinguish between two
information carrying quantum states even if they have become nonorthogonal
after passing through a channel. Therefore, quantum processes capable of
distinguishing between nonorthogonal quantum states in an optimal way offer
interesting perspectives for applications in quantum information science.
Prominent examples of such processes are the Helstrom measurement
\cite{Helstrom}, which minimizes errors, and the Ivanovic-Dieks-Peres
measurement \cite{Ivanovic,Dieks,Peres}, which distinguishes pure quantum
states in an unambiguous way.

Alternatively, nonorthogonal quantum states can also be distinguished with
the help of nonlinear quantum state transformations \cite{Gisin}. Quantum
state purification protocols
\cite{Bennett97,Deutsch98,Macchiavello98,Alber2001,Torres2016b} are early examples of
such nonlinear quantum state transformations. Thereby, identically prepared
quantum systems are subjected to an entangling unitary transformation and a
subsequent selective measurement performed on parts of the system. Iterating
these operations typically results in a strong dependence of the final state
on the initial conditions and in measurement-induced complex chaos
\cite{Kiss2006,Kiss2011}. Recently, it has been demonstrated
\cite{Gilyen2016} that the resulting strong sensitivity to initial conditions
can in principle be used to amplify small initial differences of quantum
states thus realizing a Schr\"odinger microscope, a term originally suggested
by Lloyd and Slotine \cite{Lloyd2000} capable of distinguishing non
orthogonal quantum states. Although Helstrom and Ivanov-Dieks-Peres
measurements have already been realized experimentally both optically
\cite{Clarke, Dada} and in solid state \cite{Waldherr} a Schr\"odinger
microscope based on nonlinear quantum state transformations has not yet been
realized.

Motivated by these developments the purpose of this paper is twofold, namely
to propose an experimental scenario in which iterated nonlinear dynamics can
be realized with atomic qubits and to explore the characteristic features of
the underlying nonlinear quantum state transformation in order to present a
Schr\"odinger microscope and to demonstrate its applicability for quantum
state discrimination. In view of its possibilities to measure and control the
interaction of individual atoms with a single mode of the quantized radiation
field with high precision, the area of cavity quantum electrodynamics offers
interesting perspectives for future experimental implementations in this
direction \cite{Reimann2015,Neuzner2016}. Inspired by recent experimental
advances which realize the Tavis-Cummings model \cite{Neuzner2016} in our
proposal an ensemble of identically prepared two-level atoms (qubits) is
considered which interact pairwise with a single mode of the radiation field.
Afterwards, one member of each pair and the corresponding cavity field are
measured. Conditioned on these measurement results the unmeasured atoms are
kept or discarded. In practice, this may be implemented with the help of a
single cavity and a pair of optical conveyor belts \cite{Reimann2015}, for
example. Subsequently, the atoms are moved through the cavity by the conveyor
belts in such a way that only one pair of atoms interacts with the cavity
mode at a time and then the cavity is re-initialized after each interaction.
The remaining atoms form a new identically prepared ensemble of smaller size.
Similar as in entanglement distillation protocols, the state changes of the
remaining two-level atoms are described by an iterated nonlinear quantum
transformation.

We analyze the emerging nonlinear quantum state transformation and show that
it exhibits measurement-induced complex chaos. We characterize the different
parameter regimes, the possible stable fixed points and fixed cycles of the
dynamics and the regions of convergence as well as non-converging sets of
initial states, forming the so called Julia set. Based on this analysis we
identify a case where the two stable fixed points correspond to orthogonal
quantum states of the atom and the Julia set forms a line, separating the two
regions of stability. Here the system can be utilized as a Schr\"odinger
microscope capable of amplifying the distinguishability of nonorthogonal
quantum states. In the presented setting, two-level atoms with small
excitation amplitudes can be discriminated according to the sign of the real
part of their excitation amplitudes. Thus it is also suitable to discriminate
noisy nonorthogonal quantum states.
 
This paper is organized as follows. In Sec. \ref{Model} the dynamical
equations of the two-atom Tavis-Cummings model are solved and exact and
approximate analytical solutions are presented facilitating our subsequent
treatment. Furthermore, the atomic postselection scheme is discussed which is
used in Sec. \ref{protocol} to discuss our protocol for implementing a
nonlinear map of atomic probability amplitudes. In Sec. \ref{analysis} the
fractal structure of the resulting nonlinear map is analyzed. In Sec. V our
proposal for implementing a Schr\"odinger microscope is presented. Finally,
in Sec. VI some aspects concerning possible experimental realizations of our
proposal by nowadays technology are discussed.

\section{The two-atom Tavis-Cummings model}
\label{Model}
The two-atom Tavis-Cummings model describes the resonant interaction between two atoms, say $A$ and $B$, and
a single mode of the radiation field \cite{Tavis}.
The atoms have ground states $\ket{0}_i$ and excited states 
$\ket{1}_i$ ($i \in\{A,B\}$) separated by an energy difference of $\hbar \omega$ which matches the energy of
a photon inside the empty cavity.
In the interaction picture the Hamiltonian  can be expressed in the following form
\begin{align}
  \hat{H}&= 
\hbar g\sum_{i=A,B}\left( 
\hat{\sigma}^+_i\hat{a}+  
  \hat{\sigma}^-_i\hat{a}^\dagger\right)
\label{Hamilton}
\end{align}
where 
$\hat{\sigma}^+_i=\ket{1}\bra{0}_i$ and $\hat{\sigma}^-_i=\ket{0}\bra{1}_i$
are the atomic raising and lowering operators
($i \in\{A,B\}$), 
and $\hat{a}$ ($\hat{a}^\dagger$) is the annihilation (creation) operator  of the single-mode 
field. The interaction picture is taken with respect to the reference Hamiltonian 
\begin{equation}
\hat{H}_0=\hbar\omega (\hat{a}^\dagger \hat{a}+\ketbra{1}{1}_A+\ketbra{1}{1}_B) 
  \label{H0}
\end{equation}
which is a constant of motion
as it commutes with the interaction Hamiltonian $\hat H$. For this reason both operators, $\hat H$ and $\hat H_0$,
can be diagonalized simultaneously. In fact, there is a set of common eigenvectors with eigenvalue zero, 
namely $\{\ket{\Psi^-}\ket n\}_{n=0}^\infty$.  
These states are written in terms of the Fock states $\ket n$ of the field and the atomic
states $\ket{i,j}=\ket{i}_A\ket j_B$ ($i,j \in \{0,1\}$) together with
the atomic Bell states
\begin{align}
  \ket{\Psi^\pm}&=\frac{1}{\sqrt2}\left(\ket{0,1}\pm\ket{1,0}\right).
  \label{bellstates}
\end{align}
The evaluation of the rest of the eigenvectors can be simplified by realizing that
%,as a consequence of the degenerate spectrum of $\hat H_0$, 
$\hat H$ has a block-diagonal form in the basis
\begin{align}
  &
  \{\ket{0,0}\ket0\}\oplus
  \{\ket{\Psi^+}\ket{0},\ket{0,0}\ket{1}\}\oplus
  \nonumber\\&
  \{\ket{1,1}\ket{n-2},
  \ket{\Psi^+}\ket{n-1}
  ,\ket{0,0}\ket{n}\}_{n=2}^\infty,
  \label{basisAB}
\end{align}
with blocks given by the following matrices
\begin{align}
  &H^{(0)}=0,\quad
 H^{(1)}=\hbar g\left(
  \begin{array}{cc}
    0& \sqrt{2}\\
    \sqrt{2}& 0
  \end{array}
 \right),
  \nonumber\\
  &H^{(n\ge 2)}=
  \hbar g\left(
  \begin{array}{ccc}
   0 &\sqrt{2(n-1)}&0\\
     \sqrt{2(n-1)}&0&\sqrt{2n}\\
    0&\sqrt{2n}& 0
  \end{array}
  \right).
  \label{Hblocks1}
\end{align}
The eigenvalues of these matrices are given by $\{0\}$ for $n=0$, 
$\{ -\sqrt2\hbar g,\sqrt2\hbar g\}$ for $n=1$, and 
$\{0,-\hbar\omega_n,\hbar\omega_n\}$ for $n\ge 2$, with
\begin{align}
  \omega_n=g\sqrt{4n-2}.
  \label{}
\end{align}
The transformations that diagonalize each of the blocks  $H^{(n)}$ are given by
\begin{align}
  O^{(1)}&=\frac{1}{\sqrt2}\left(
 \begin{array}{cc}
   1&1\\
   -1&1
 \end{array}
  \right),
  \\
  O^{(n\ge2)}&=\frac{1}{\sqrt{4n-2}}\left(
  \begin{array}{ccc}
    -\sqrt{2n}&\sqrt{n-1}&\sqrt{n-1}\\
    0&-\sqrt{2n-1}&\sqrt{2n-1}\\
    \sqrt{2n-2}&\sqrt n&\sqrt n
  \end{array}
  \right).\nonumber
  \label{}
\end{align}
These matrices  are the blocks of the orthogonal transformation $\hat O$ that diagonalizes the Hamiltonian 
$\hat H$ as $\hat O^\dagger \hat H \hat O$. 

\subsection{Exact solution}
Having solved the eigenvalue problem for $\hat H$, it is now possible to evaluate the time-dependent
state vector 
\begin{align}
  \ket{\Psi_t}= e^{-i\hat H t/\hbar}\ket{\Psi_0}
  \label{}
\end{align}
for any given initial pure state $\ket{\Psi_0}$. In this work we 
consider as initial condition a normalized product state of the two atoms and the single-mode field that can be expressed as
\begin{align}
  \ket{\Psi_0}=&\ket{\Psi_0^{\rm at}}\ket\alpha,
  \nonumber\\
  \ket{\Psi^{\rm at}_0}=&
  c_0\ket{0,0}+
  c_-\ket{\Psi^-}+
  c_+\ket{\Psi^+}+
  c_1\ket{1,1}.
  \label{initial}
\end{align}
We have considered a general pure state
$\ket{\Psi^{\rm at}_0}$ of the atoms
with probability amplitudes $c_\pm$, $c_0$ and $c_1$. For the single mode of the radiation field 
we have chosen a coherent state
\begin{align}
  \ket{\alpha}=\sum_{n=0}^\infty 
  e^{-\frac{|\alpha|^2}{2}}
 \frac{\alpha^n}{\sqrt{n!}}
 \ket n,
  \quad\alpha=\sqrt{\overline n}\,e^{i\phi},
  \label{coherentstate}
\end{align}
with mean photon number $\bar n$. Using the eigenbasis of $\hat H$,  
the exact solution of the time-dependent state vector can be written as
\begin{equation}
  \ket{\Psi_t}=
  \ket{0,0}\ket{\chi_t^{-1}}
  +\ket{\Psi^+}\ket{\chi^0_t} 
  +\ket{1,1}\ket{\chi^{1}_t}
  +c_-\ket{\Psi^-}\ket{\alpha}
  \label{psi}
\end{equation}
with the relevant photonic states
\begin{align}
  \ket{\chi^{-1}_t}&=
  c_0\,p_0\ket{0}+
  \sum_{n=1}^\infty 
  \frac{
  \sqrt{n}
  \left(
  \xi_{n,t}^-
  -\xi_{n,t}^+\right)
  +\sqrt{n-1}\xi_{n}
  }{\sqrt{2n-1}}
  \ket{n},
  \nonumber\\
  \ket{\chi^0_t}&=
  \sum_{n=1}^\infty 
  \left(
  \xi_{n,t}^-+\xi_{n,t}^+
  \right)
  \ket{n-1},
  \label{fieldstates}
  \\
  \ket{\chi^1_t}&=\sum_{n=2}^\infty 
  \frac{
  \sqrt{n-1}
  \left(
  \xi_{n,t}^-
  -\xi_{n,t}^+\right)
  -\sqrt{n}\xi_{n}
  }{\sqrt{2n-1}}
  \ket{n-2},
  \nonumber
\end{align}
and with the aid of the following abbreviations
\begin{align}
  \label{abbr}
  &\xi_{n,t}^\pm
  =
  \frac{e^{\pm i \omega_n t}}{2}
  \left(
  c_+\mp
  \frac{\,c_0p_n+\sqrt{n-1}\,c_1p_{n-2}}{\sqrt{2n-1}}
  \right),
  \\
  &\xi_{n}
  =
  \frac{\sqrt{n-1}\,c_0 p_n-\sqrt{n}\,c_1 p_{n-2}}{\sqrt{2n-1}},\quad
  p_n=\alpha^n\sqrt{e^{-|\alpha|^2}/n!}.
  \nonumber
\end{align}

\subsection{Coherent-state approximation}
The time-dependent solution of the state vector can be significantly simplified
in the case of high values of the mean photon number, i.e., $\bar n\gg 1$. In this limit 
 $\bar n\gg \sqrt{\bar n}$, i.e., the mean of the Poisson distribution $\bar n$ 
is much larger than the standard deviation $\sqrt{\bar n}$. Therefore we approximate
$\sqrt{(n-1)/(2n-1)}$
and $\sqrt{n/(2n-1)}$ by $1/\sqrt2$ and we also use the approximations
\begin{align}
  p_n=&\sqrt{\frac{\bar n}{n}}e^{i\phi}p_{n-1}\approx e^{i\phi}p_{n-1},
  \nonumber\\
  \omega_n/g&\approx\sqrt{4 \bar n+2}+2\frac{n-\bar n-1}{\sqrt{4\bar n+2}}.
  \label{apps}
\end{align}
The last line is obtained from the first-order Taylor expansion in $n$ of the frequencies around $\bar n+1$.
This is valid whenever the product between the second-order contribution times the interaction time $t$ remains small,
a condition that is satisfied when $gt\ll\bar n$ \cite{Torres2014,Torres2016}.
With these considerations and by introducing the abbreviations
\begin{align}
  \eta_\pm=\frac{1}{2}\left(
  c_+\mp d^+_\phi
  \right),\quad d_\phi^\pm=
  \frac{e^{i\phi} c_0\pm e^{-i\phi}c_1}{\sqrt2},
  \label{}
\end{align}
the photonic states can be simplified to
\begin{align}
  \ket{\chi^k_t}&\approx
  \frac{
  e^{ik\phi}}{\sqrt{1+|k|}}
  \left(
  \eta_-
  \ket{F_{k,t}^-}
  +(-1)^k\eta_+
  \ket{F_{k,t}^+}
  -kd_\phi^-\ket\alpha
  \right),
  \nonumber\\
   k&\in\{-1,0,1\},
  \label{fieldstates2}
\end{align}
where we have introduced the field states 
\begin{align}
  \ket{F_{k,t}^\pm}=e^{\pm i 2gt\frac{1+k(\bar n+1)}{\sqrt{4 \bar n+1}}}\ket{\alpha e^{\frac{\pm i2gt}{\sqrt{4\bar n+1}}}},
  \quad k\in\{-1,0,1\}.
  \label{Fstates}
\end{align}
which are coherent states up to an additional phase. 

\subsection{Atomic postselection}
The description in terms of coherent states allows a simpler
analysis of the dynamics. Our aim is to prepare the atoms in  an atomic postselection scenario where the atoms are prepared conditioned
to a successful projection of the field onto the initial coherent state $\ket\alpha$ in a simplified and ideal 
implementation. 
In such a case, one would have to consider the following overlaps
\begin{align}
  \braket{\alpha}{\chi_t^{k}}\approx-k
  e^{ik\phi}
  \frac{e^{i\phi} c_0\pm e^{-i\phi}c_1}{2}.
  \label{18}
\end{align}
This result can be obtained by noting that the overlap between coherent states is given by
\begin{align}
  \left|\braket{\alpha}{\alpha e^{\frac{\pm i2gt}{\sqrt{4\bar n+1}}}}\right|
  =&\left|\exp{\left[-\bar n\left(1-e^{\frac{\pm i2gt}{\sqrt{4\bar n+1}}}\right)\right]}\right|
  \approx&e^{- g^2t^2},
  \nonumber
  \label{}
\end{align}
which can be neglected if $ gt\gg1$. Therefore, after the interaction with the resonator and 
projection onto state $\ket\alpha$, both atoms are left in the state
\begin{align}
  \frac{c_-}{Q_1}\ket{\Psi^-}+
  \frac{e^{i\phi} c_0- e^{-i\phi}c_1}{2 Q_1}
  \left(
  e^{-i\phi}\ket{0,0}-
  e^{i\phi}\ket{1,1}
  \right),
  \label{atpost}
\end{align}
with $Q_1^2=|c_1|^2+|e^{i\phi} c_0- e^{-i\phi}c_1|^2/2$  success probability. The final state
is actually a superposition of two states with probability amplitudes proportional to the intial ones.
Therefore, the atomic postselection can be understood as a projection of the atomic state with
the following rank two projector
\begin{equation}
  \hat M= 
  \ketbra{\Psi^-}{\Psi^-}+\ketbra{\Phi^-_\phi}{\Phi^-_\phi},
  \label{Moperator}
\end{equation}
where we have introduced the state $\ket{\Phi^-_\phi}=\left(
  e^{-i\phi}\ket{0,0}-  e^{i\phi}\ket{1,1} \right)/\sqrt2$. The operation $\hat M$ represents the effective
 description of the interaction of the atoms with the resonator and the postselection via measurement of the field.

\subsection{Atomic postselection by balanced homodyne detection}
Considering the projection onto a coherent state is an idealization that provides a convenient simplified picture.
In practice, however, it is sufficient to project onto a state with vanishing overlap with the time-dependent field
components $\ket{F_{k,t}^\pm}$ and with finite overlap with $\ket\alpha$. 
A typical experimental setting able to achieve this goal is a  balanced homodyne measurement \cite{Torres2014}.
The basic idea is to use a $50/50$ beam splitter to combine
the field to be measured with a reference coherent field parametrized by
its phase $\theta$. Photons from the two outputs of the beam splitter are collected using photodetectors.
In the strong limit of the reference field  and assuming ideal photodetectors \cite{Raymer}, 
the probability of measuring a photocurrent difference  between the detectors is proportional to the projection of
the field onto the eigenstate $\ket{q_{\theta}}$ of a field quadrature 
$\hat q_\theta=(\hat a e^{-i\theta}+\hat a^\dagger e^{i\theta})/\sqrt2$. This probability density for a coherent
state $\ket\alpha$ is given by 
\begin{equation}
  |\braket{q_\theta}{\alpha}|^2=\frac{1}{\sqrt{\pi}}
  \exp\left\{-[ q_\theta-\tilde q_\theta]^2\right\}
  \label{probcoh1}
\end{equation}
with $\tilde q_\theta =(\alpha e^{-i\theta}+\alpha^\ast e^{i\theta})/\sqrt2$. This overlap can approach 
its maximum value by choosing the phase in such a way that $\tilde q_\theta=0$ and restricting values of
$q_\theta$ close to zero.
The square of the overlap with the
other field components, that are also coherent states, can be evaluated as 
\begin{equation}
  |\braket{q_\theta}{F_{kmt}^\pm}|^2=\frac{1}{\sqrt{\pi}}
  \exp\left\{-[ q_\theta-\tilde q_{\Theta_t^\pm}]^2\right\},
  \label{probcoh2}
\end{equation}
with $\Theta_t^\pm=\theta\mp 2gt/\sqrt{4\bar n +1}$. By choosing an appropriate interaction time $t$, these
overlaps can be made exponentially small.

\section{A nonlinear map of pure atomic states}
\label{protocol}

%\subsection{Single implementation of the map}
\label{onestep}
\begin{figure}[t]
\includegraphics[width=0.47\textwidth]{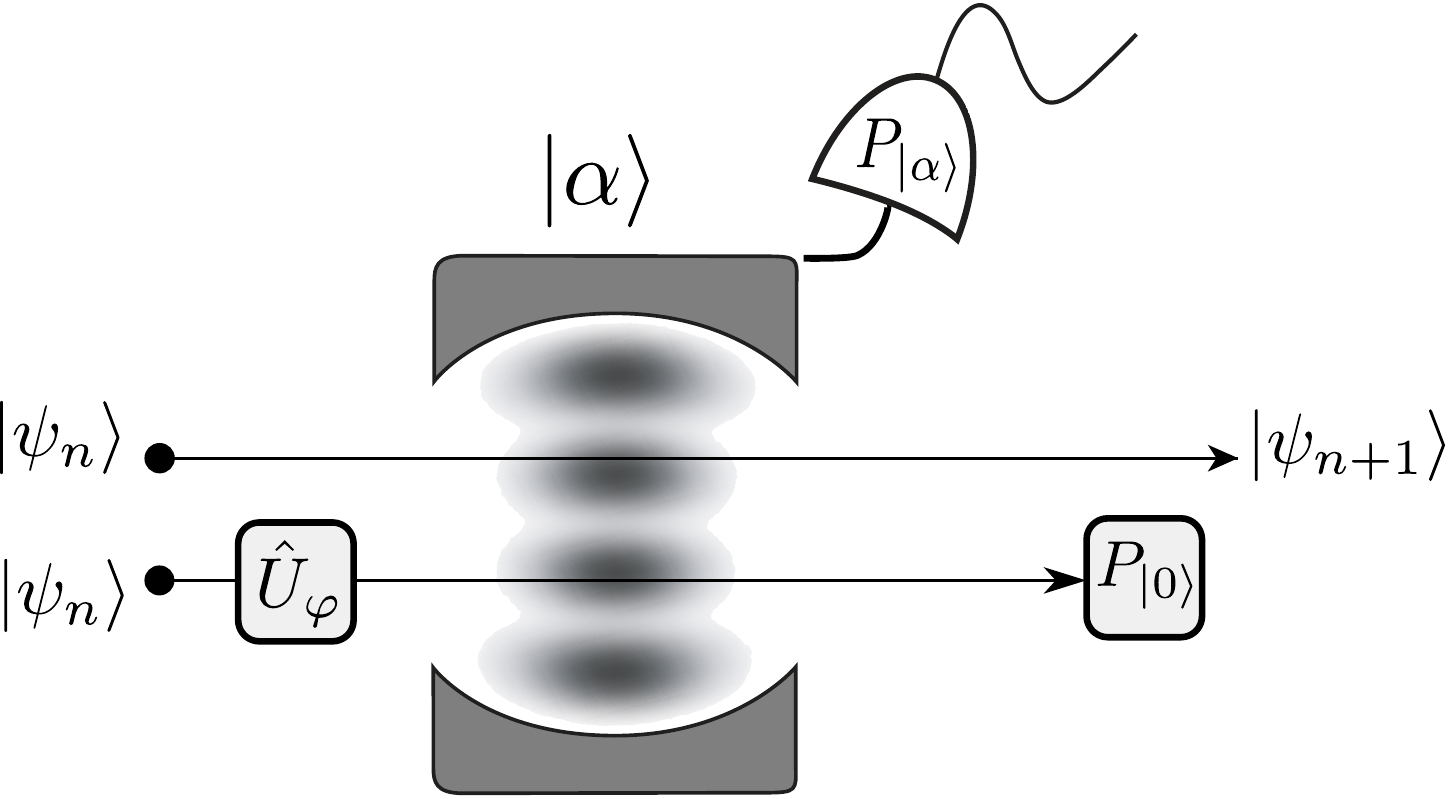}
\caption{\label{scheme}
Two two-level atoms in the same state $\ket{\psi_n}$
interact with the cavity field prepared in a coherent state $\ket{\alpha}$.
Before the interaction, the gate $\hat U_\varphi$ is applied to one of the atoms
and after the interaction and the projection of the field onto the initial coherent state, this same atom is projected onto its ground state. Finally, the other atom is left in the state $\ket{\psi_{n+1}}$.
}
\end{figure}

In this section we use the atomic postselection scheme of the two-atom Tavis-Cummings model in order to implement
an entangling quantum operation which by iteration leads to a nonlinear
mapping of atomic probability amplitudes. 
The protocol is depicted schematically in Fig. \ref{scheme}.
We consider the two two-level atoms
initially prepared in a product state of the form ($z\in\mathbb{C}$)
\begin{align}
  \ket{\Psi^{\rm at}_0}=\ket{\psi_0}_A\otimes\ket{\psi_0}_B,  \quad
  \ket{\psi_0}=\frac{\ket{0}+ze^{i\phi}\ket{1}}{\sqrt{1+|z|^2}}.
  \label{atom-state}
\end{align}
For later convenience we have included the phase $\phi$ of the coherent state.
Before interacting with the optical resonator, a unitary gate $\hat U^B_\varphi$ is applied to atom $B$. 
We choose the following gate 
\begin{align}
  \hat U_\varphi=\left(
  \begin{array}{cc}
    e^{i\varphi}&0\\
    0&-e^{-i\varphi}
  \end{array}
  \right),
  \label{gate}
\end{align}
which can be implemented by driving the atomic transition with a resonant classical electromagnetic field and properly
controlling the coupling and duration of the interaction \cite{Nielsen,Meschede2006,Raimond}.
After the application of $\hat U_\varphi^B$ and 
before entering the resonator we get the following atomic probability amplitudes
\begin{align}
  c_0&=\bra{0,0}\hat U^B_\varphi\ket{\Psi^{\rm at}_0}=-e^{-i\varphi}/(1+|z|^2)
  \nonumber\\
  c_1&=\bra{1,1}\hat U^B_\varphi\ket{\Psi^{\rm at}_0}=z^2e^{i(\varphi-2\phi)}/(1+|z|^2)
  \nonumber\\
  c_-&=\bra{\Psi^-}\hat U^B_\varphi\ket{\Psi^{\rm at}_0}=\sqrt2 z e^{i\phi} \cos \varphi/(1+|z|^2).
  \label{}
\end{align}
The probability amplitude $c_+$ does not need to be specified, as the resulting
quantum operation projects the atoms onto a subspace orthogonal to $\ket{\Psi^+}$
as can be noted from Eq. \eqref{atpost}.
With these initial conditions, both atoms interact with
the electromagnetic field inside a cavity prepared in a coherent state $\ket\alpha$.
After the interaction  a projection $P_{\ket\alpha}$ of the field onto the initial coherent
state $\ket\alpha$ is performed and the atoms are left in the state
\begin{align*}
  &\frac{\sqrt2ze^{i\phi}\cos\varphi}{(1+|z|^2)Q_1}\ket{\Psi^-}
  -\frac{ e^{-i\varphi}+z^2 e^{i\varphi}}{2(1+|z|^2)Q_1}
  \left(\ket{0,0}-e^{i2\phi}\ket{1,1}\right).
\end{align*}
The success probability of this projection is 
\begin{equation}
  Q_1^2=\frac{1+|z|^4+4|z|^2\cos^2\varphi+(z^2e^{i2\varphi}+{\rm c.c.})}{2(1+|z|^2)^2}.
  \label{Q1}
\end{equation}
Afterwards a projection $P_{\ket{0}}$ onto the ground state of  atom $B$ is implemented
leaving atom $A$  in the state
\begin{align}
  &-\frac{ze^{i\phi}\cos\varphi}{(1+|z|^2)Q_1Q_2}\ket{1}
  -\frac{ e^{-i\varphi}+z^2 e^{i\varphi}}{2(1+|z|^2)Q_1Q_2}
  \ket{0}.
  \label{}
\end{align}
This event occurs with success probability
  $Q_2^2=1/2$.
The overall success probability of the postselections is then given by
\begin{equation}
  P_{\rm s}=Q_2^2Q_1^2=Q_1^2/2\ge \frac{\cos^2\varphi}{4}.
  \label{psuccess}
\end{equation}
The last inequality follows from analyzing Eq. \eqref{Q1} and noting that $Q_1$
attains its minimum value when $|z|^2=1$ and ${\rm Re} [z^2e^{i2\varphi}]=-1$.
Up to normalization the final state is given  by
\begin{align}
  \ket{0}+\frac{2 z\cos\varphi}{e^{-i\varphi}+z^2 e^{i\varphi}}e^{i\phi}\ket{1}.
  \label{}
\end{align}
By iterating this procedure we attain a scheme implementing the following quantum map
for the $(n+1)$th step
\begin{align}
  \frac{\ket{0}+f^{ n}_\varphi(z)e^{i\phi}\ket{1}}{\sqrt{1+|f_\varphi^{n}(z)|^2}}
  \rightarrow
  \frac{\ket{0}+f^{ n+1}_\varphi(z)e^{i\phi}\ket{1}}{\sqrt{1+|f_\varphi^{n+1}(z)|^2}}
  \label{nmap}
\end{align}
with the complex functions
\begin{align}
&f_\varphi(z)=\frac{ 2z\cos\varphi}{e^{-i\varphi}+z^2 e^{i\varphi}},
\nonumber\\
&f^{n+1}_\varphi(z)=f_\varphi(f_\varphi^n(z)),\quad f^0_\varphi(z)=z.
\label{map}
\end{align}
The map is independent of the parameter $\phi$, as 
one can note that the phase factor $e^{i\phi}$ appears in the probability amplitude
of state $\ket{1}$ in the same manner as in the initial state $\ket{\psi_0}$ of Eq. 
\eqref{atom-state}.

We note that the iteration of the map involves repeated action of the protocol on an ensemble of atoms. The protocol acts on a pair of identically prepared atoms from the ensemble and prepares one atom probabilistically. The other atom becomes useless from the point of view of the protocol, as a result of the projective measurement on it. After acting on all the atoms of the ensemble, one arrives at a smaller ensemble of less than one half in size. Rapid downscaling of the ensemble size is a unavoidable condition for any quantum dynamics truly sensitive to initial conditions \cite{Gilyen2016}. 
In practice, realizing many steps of the protocol would require an exponentially large initial ensemble which would not be realistic. Another practical aspect is that employing more than one cavity would be challenging with today's experimental possibilities. On the other hand, as we will demonstrate in the next sections, already a few steps can be enough to make highly overlapping initial quantum states almost orthogonal.  Furthermore we will outline an experimental proposal in Sec. \ref{Section-experimental} with currently available technology by applying an optical conveyor belt and a single cavity.

\section{Basic properties of the nonlinear map}
\label{analysis}

The dynamics within the approximations we have made is fully described by the iterative complex function in Eq. (\ref{map}). This is a quadratic rational map \cite{MilnorGD}, similar to the maps occurring in the measurement-induced nonlinear quantum dynamical schemes first described in \cite{Gisin,Kiss2006,Gilyen2016}. In the following, we first carry out an  analysis of the general properties of the iterated map $f_{\varphi}$ of Eq. (\ref{map}) by using concepts from the theory of complex dynamical maps \cite{Milnorbook}. Then we compare its behavior to the numerical solution of the complete iterated dynamics, based on the Hamiltonian of Eq. (\ref{Hamilton}) and the subsequent selective measurements.

\subsection{Stable cycles}
\begin{figure}[t]
\includegraphics[width=1.0\columnwidth]{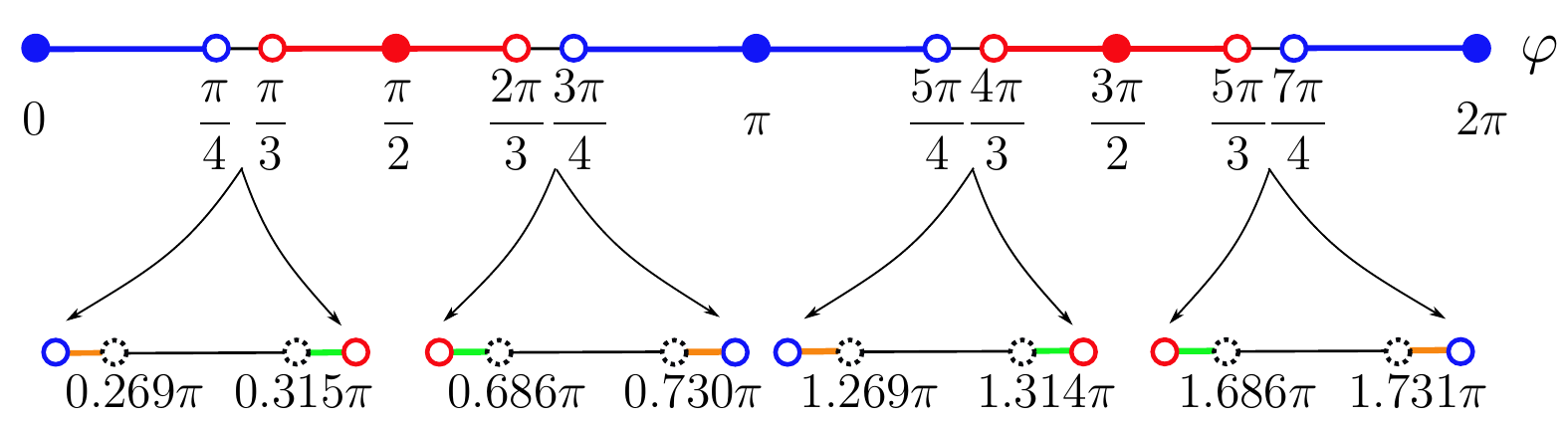}
\caption{Stability of the fixed cycles of $f_{\varphi}$ as a function the parameter $\varphi$. Blue corresponds to the one-cycles $z^{(1)}=\pm 1$, red corresponds to $z^{(1)}=0$. Dots, lines, and circles represent superattractive, attractive, and neutral cycles, respectively. Numerical investigation of the enlarged regions between the neutral one-cycles show two different attractive 4-cycles (orange lines) and a single attractive 6-cycle (green lines) close to the two ends of the region. The central part of the enlarged regions contain "islands" of attractive $n\geq 60$ cycles. The dotted circles indicate that is hard to identify the border of different regions.}
\label{Fig2}
\end{figure}

The periodic orbits or fixed cycles of the map $f_{\varphi}$ can be determined from the relation $f^{n}_{\varphi}(z)=z$. The one-cycles or fixed points as well as the 2-cycles can be determined analytically.
For $n=1$ we find
\begin{equation}
z^{(1)}_{j}=j, \quad j\in\{-1,0,1\}.
\end{equation} 
For $n=2$, in addition to the above one-cycles, one can find two more points which are transformed into each other by $f_{\varphi}$. These form the single nontrivial two-cycle 
\begin{equation}
z^{(2)}_{k}=(-1)^{k}i\sqrt{1+2e^{-2i\varphi}}, \quad k\in\{1,2\}.
\label{2cyc}
\end{equation}

The stability of the fixed cycles can be checked by calculating the multiplier $\lambda=\left(f^{n}_\varphi\right)\rq(z_{j})= f\rq_{\varphi}(z_{1})f\rq_{\varphi}(z_{2})...f\rq_{\varphi}(z_{n})$. A fixed cycle is repelling, neutral, attractive, or superattractive  if $\left| \lambda \right|>1$, $\left| \lambda \right|=1$, $\left| \lambda \right|<1$, or $\left| \lambda \right|=0$, respectively. Such an analysis can be carried out analytically for the one- and two-cycles, however, for $n\geq 2$ it is a nontrivial task. The analysis of the multipliers shows that for each of the one-cycles there are certain parameter regions where they are attractive. On the other hand, the two-cycle given by Eq. (\ref{2cyc}), is repelling for any value of $\varphi$.

For the determination of the longer ($n\geq$3) attractive cycles we can use the method based on the iteration of the critical points of the map. The critical points of $f_{\varphi}$ are those which solve the equation $f\rq_{\varphi}(z)=0$. In this case, there are two critical points:
\begin{equation}
z_{c\pm}=\pm e^{-i\varphi}.
\label{crit}
\end{equation} 
A general theorem on iterated rational polynomial maps states that a rational map of degree $d$ can have at most $2d-2$ attractive cycles. Following the orbits of the critical points one can find all stable cycles of the iterated map (in this case at most $2$).

Fig.~\ref{Fig2} shows where, according to the analytical calculations, the one-cycles are superattractive (dots), attractive (lines), and neutral (circles) as a function of the parameter $\varphi$. The numerical iteration of the critical points in the regions between the neutral one-cycles shows that there are two different attractive 4-cycles (orange lines) and a single 6-cycle (green lines) close to the two ends of the regions. The actual $z$ values belonging to the attractive 4- and 6-cycles depend on the parameter $\varphi$. In between these regions, it is numerically hard to rule out the existence of very long stable periodic orbits. The precision of our numerical simulation made it possible to identify a few "islands" of attractive fixed cycles of $n\geq 60$. The remaining part of this region may belong to maps without any stable periodic orbit, which means that all initial states belong to the Julia set. The dotted circles indicate that the border between different regions is hard to determine numerically, which is an indication of the fractal nature of the regions. Let us note that for $\varphi=\pi/2$ and $3\pi/2$ the map is actually not a genuine complex map since $f_{\varphi}\equiv 0$ in these cases.

\subsection{Nature of the iterated map}

\begin{figure}[tbh]
\includegraphics[width=0.9\columnwidth]{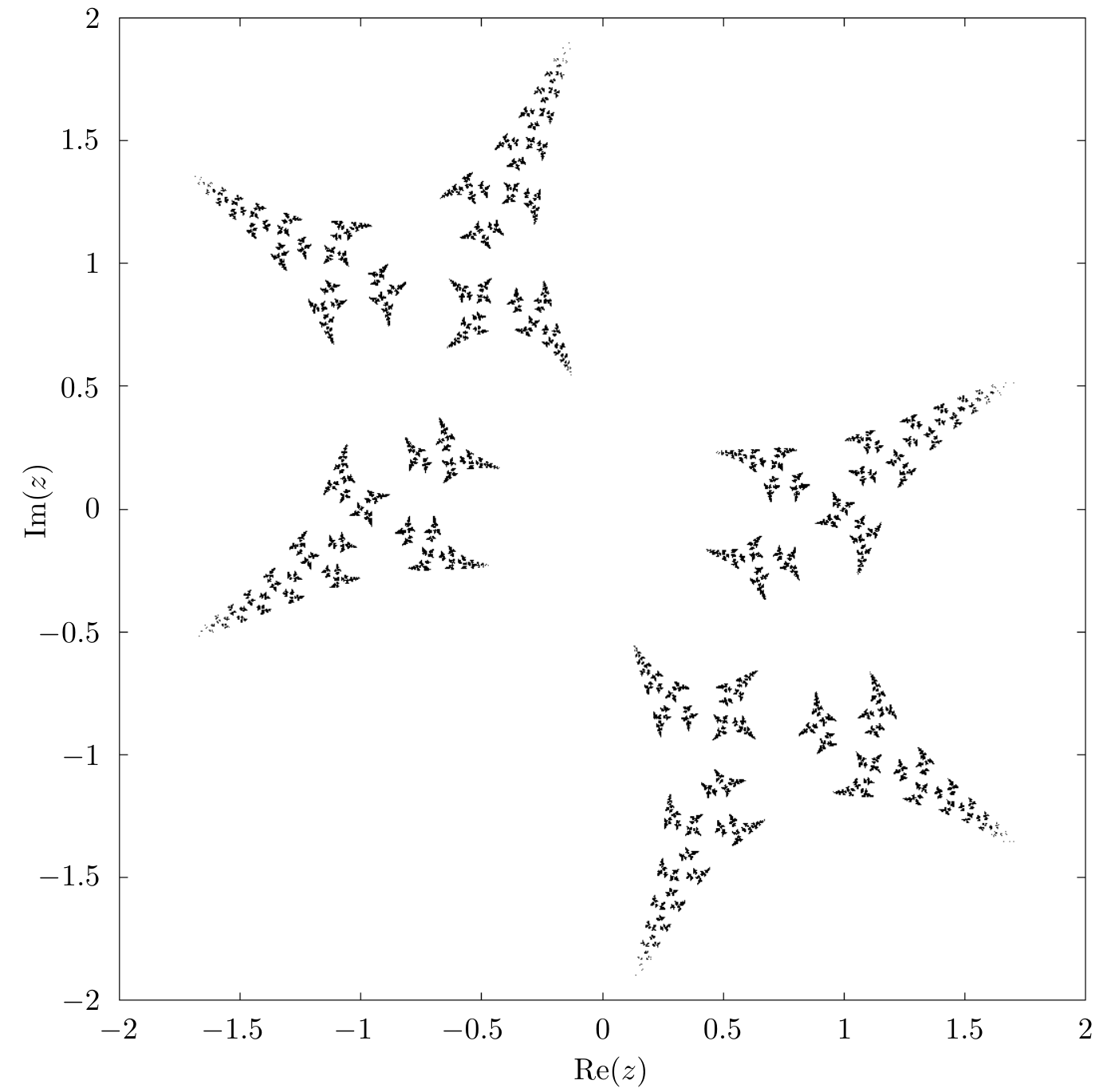}
\caption{
The Julia set of the map $f_{\varphi}$ for $\varphi=1.666\pi$.
}
\label{Fig3}
\end{figure}

\begin{figure}[t]
\includegraphics[width=0.49\columnwidth]{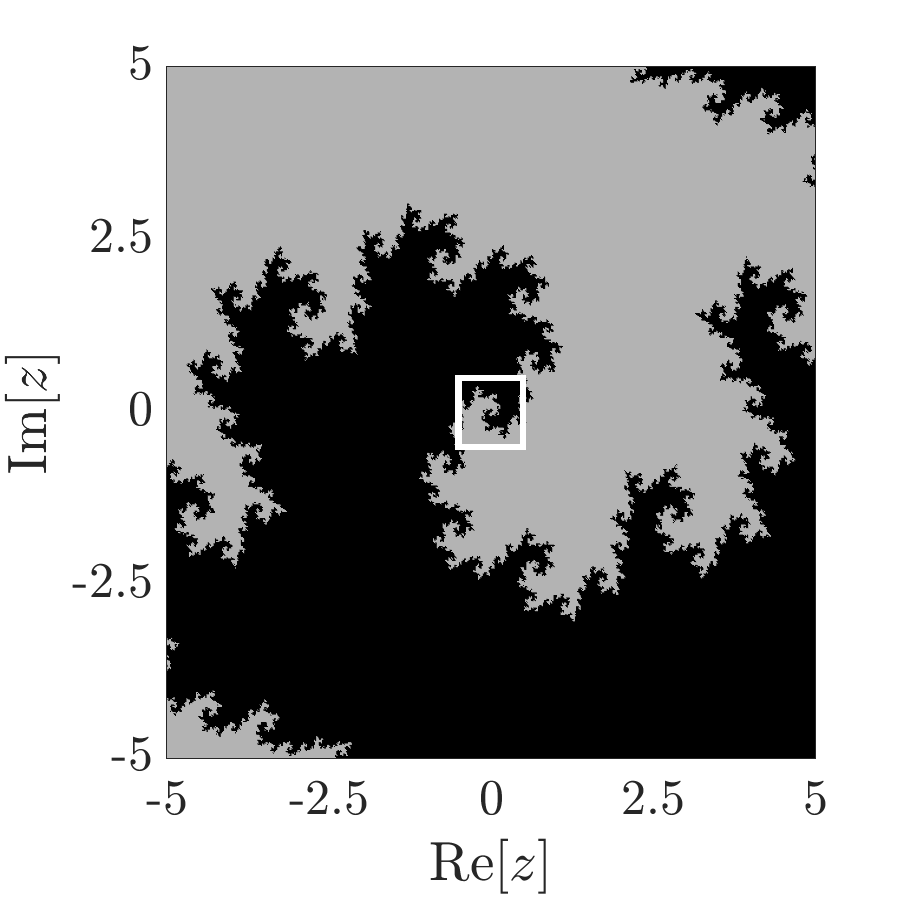}
\includegraphics[width=0.49\columnwidth]{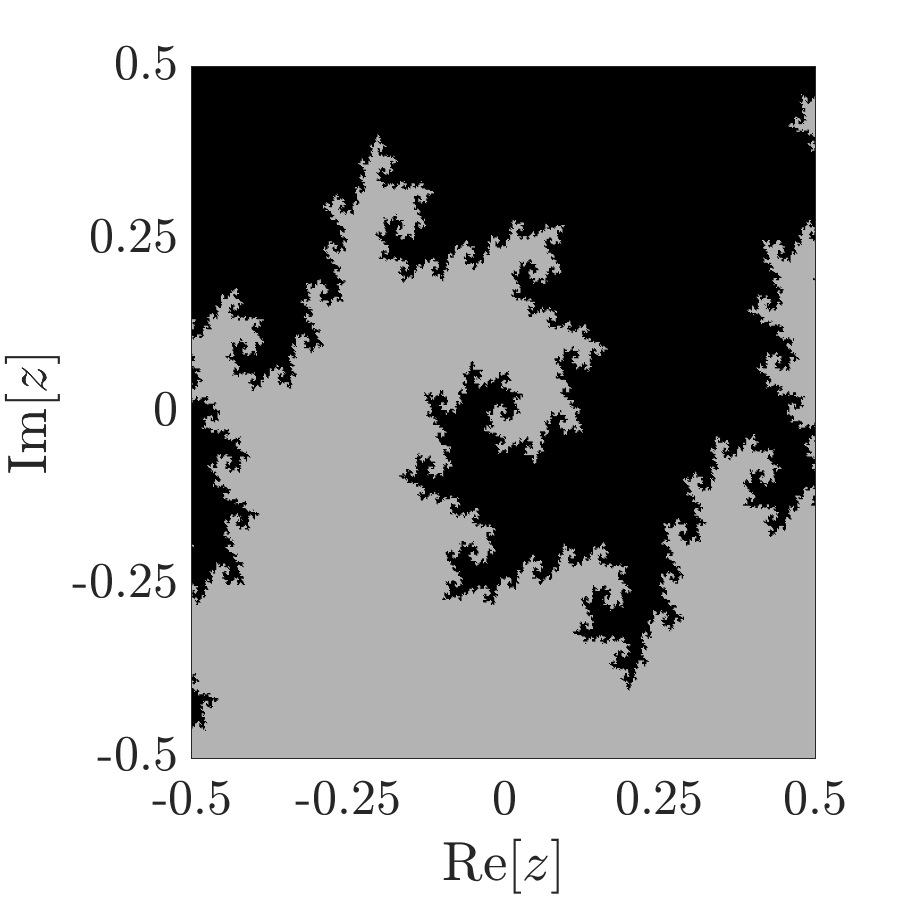}
\caption{\label{cplane1}
Complex plane after $97$ iterations of the map in Eq. \eqref{nmap} for $\varphi=0.95\pi/4$. Two amplification levels are shown,
confirming the fractal structure of the Julia set separating the regions whose points  converge to the attractive fixed points $1$ (grey) and $-1$ (black).
The region indicated by the square in the middle of the left figure is magnified in the right figure.
}
\end{figure}

The fractal nature of the map is more apparent when one determines the Julia set of $f_{\varphi}$, i.e. the set of points which do not converge to an attractive cycle for a given $\varphi$. One way of numerically finding the points belonging to the Julia set is backwards iterating the map starting from a point which is an element of a repelling cycle of the map. We show in Fig.~\ref{Fig3} the Julia set of $f_{\varphi}$ for $\varphi=1.666\pi$. In this case, the Julia set is a totally disconnected set, all other initial points converge to the single attractive cycle $z=0$, or physically speaking to the state $\ket{0}$. The analysis of the orbits of the critical points reveals important properties of the Julia set. In this case both critical points converge to the same attractive fixed point, consequently the Julia set is totally disconnected, similarly to the well-known Cantor set \cite{MilnorGD}.
Another important case is when the two critical points converge to two distinct fixed points, then the Julia set is connected. This case is illustrated by the map at parameter value $\varphi=0.95\pi/4$ shown in Fig.~\ref{cplane1}.
For quadratic rational maps a general theorem ensures that the Julia set is either totally disconnected, or connected \cite{Milnorbook}.

\subsection{Iteration of the complete dynamics}

\begin{figure}[b]
\includegraphics[width=0.49\columnwidth]{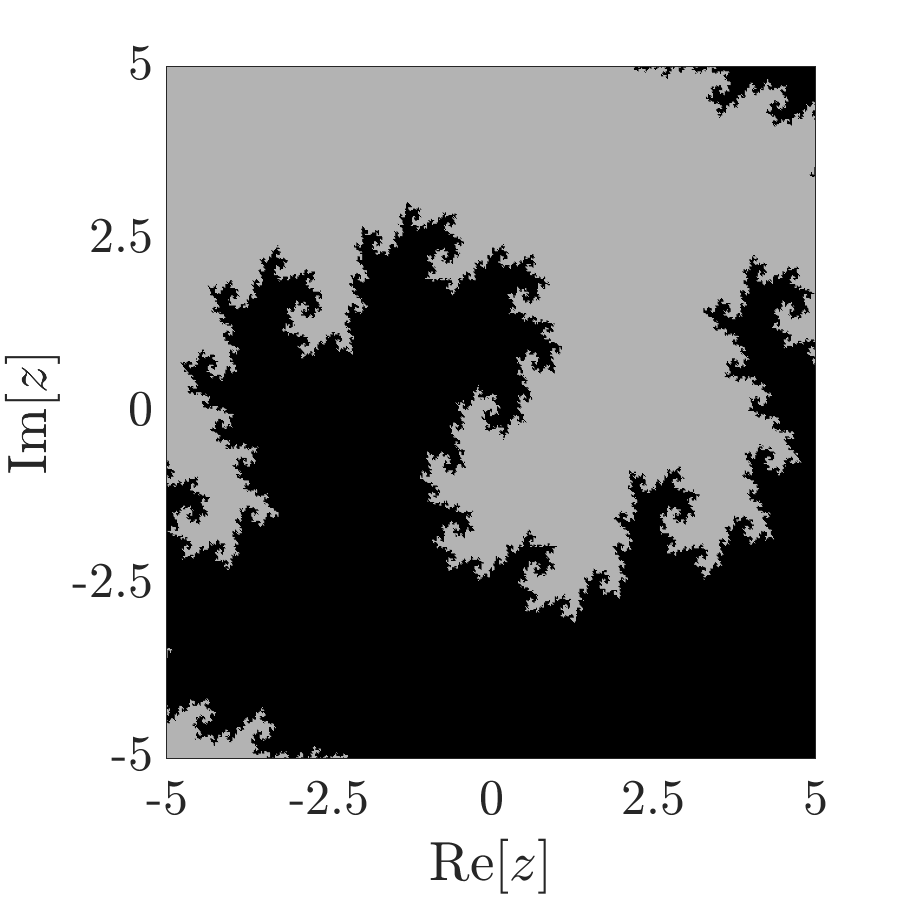}
\includegraphics[width=0.49\columnwidth]{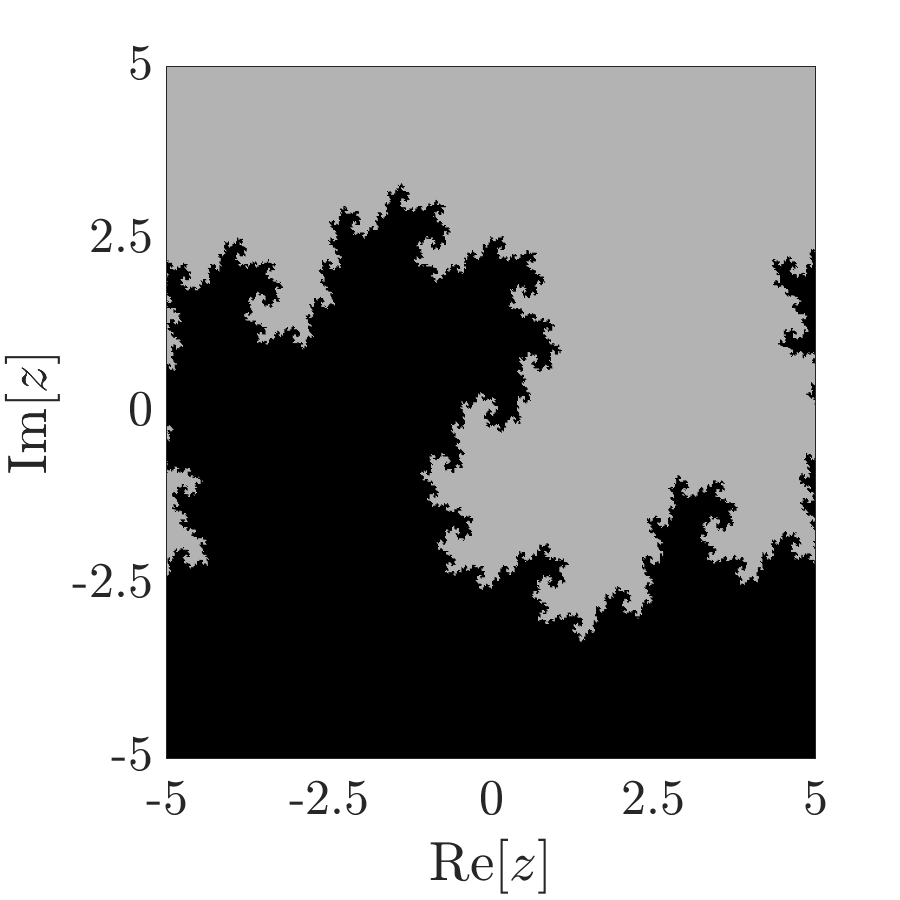}
\caption{\label{cplane2}
%Complex plane after $97$ iterations of 
The same as left part of Fig. \ref{cplane1} for the 
the numerically exact quantum map and 
two values of the mean photon number $\bar n$: $100$ (left) and $10$ (right).
}
\end{figure}

In order to investigate the real performance of the two-atom Tavis-Cummings model
without the approximations of Sec. \ref{Model}, we compute a numerically exact version of the 
operator $\hat M$ in Eq. \eqref{Moperator}. The matrix elements are evaluated as
\begin{align}
  M_{j,k}=\bra\alpha\bra{e_j}e^{-i \hat H t/\hbar} \ket{e_k}\ket\alpha
  \label{qmap}
\end{align}
where we considered the atomic basis  
$\ket{e_j}\in\{\ket{1,1},\ket{1,0},\ket{0,1},\ket{0,0}\}$. 
The interaction time $t$ and coupling strength $g$ satisfy the relation
$gt=\pi\sqrt{\bar n}/2$.
Each iteration of the map is then
evaluated by renormalizing the following outcome $\bra{0}_B\hat M\hat U_B\ket{\Psi_0^{\rm at}}$
for qubit $A$. In Fig.~\ref{cplane2} we plotted the real part of the $97$th iteration for two different values of the mean photon number $\bar n$, namely $100$, and $10$. 
With precision of two (one) decimal places the two fixed points also converge to +1 and -1
in the case of $\bar n=100$ ($\bar n=10$).
Both figures reveal a fractal structure which resembles more the ideal case for larger values of $\bar n$.

\section{Application of the protocol for state discrimination}

The number of atoms needed by a protocol based on a nonlinear transformation grows exponentially with the number of iterations even in an ideal case, which follows from the quantum magnification bound \cite{Gilyen2016}. In a realistic experiment, one can expect that only a few steps of the iteration can be carried out. On the other hand, a useful aspect of nonlinear quantum state transformations is that small initial differences between two similar quantum states can be amplified, enabling to distinguish them, realizing a Schr\"odinger microscope \cite{Lloyd2000}. Nonlinear quantum state transformations in an ideal case saturate the quantum magnification bound \cite{Gilyen2016}, thereby providing an optimal quantum state discrimination protocol, according to Helstrom \cite{Helstrom}. Here we show that our protocol provides a practical state discrimination procedure, transforming initially very close states into almost perfectly orthogonal ones in as few as 3 steps.

\begin{figure}[t]
\includegraphics[width=0.9\columnwidth]{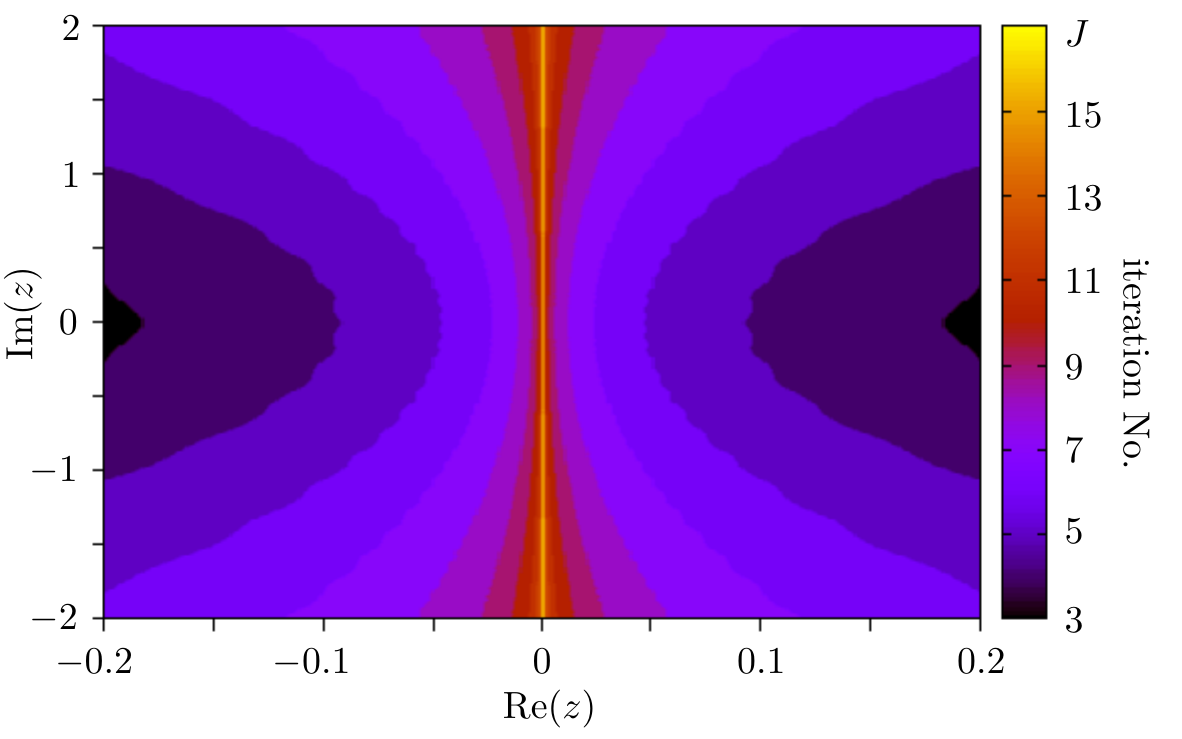}
\caption{
The plane of initial states colored according to the number of iterations needed for a complex number $z$ to reach either of the fixed points $1$ or $-1$ with a precision of $0.1$.}
\label{numit}
\end{figure}

In the simple case when $\varphi=0$, the nonlinear map reads $f_{\varphi=0}=2z/(z^{2}+1)$ and the unitary of Eq. (\ref{gate}) is the well-known Z gate. In Fig.~\ref{numit} we show the plane of initial states colored according to the number of iterations needed to reach either of the fixed points $1$ or $-1$ with a precision of $0.1$. Complex numbers with a positive (negative) real part converge to the 
fixed point $1$ ($-1$). The two regions are separated by the Julia set of the map, which is indicated by the yellow region on the figure, coinciding with the imaginary axis. If we choose two initial quantum states close to each other in the form of Eq. (\ref{atom-state}) with $z_{1}=-0.2$, and $z_{2}=0.2$ (with an overlap close to unity $|\braket{\psi^{1}_{0}}{\psi^{2}_{0}}|\sim 0.92$), then the two states will become almost orthogonal (with a scalar product of $\sim 0.08$) after three steps of the iteration. 

\begin{figure}[tbh]
\subfigure{
\includegraphics[width=0.73\columnwidth]{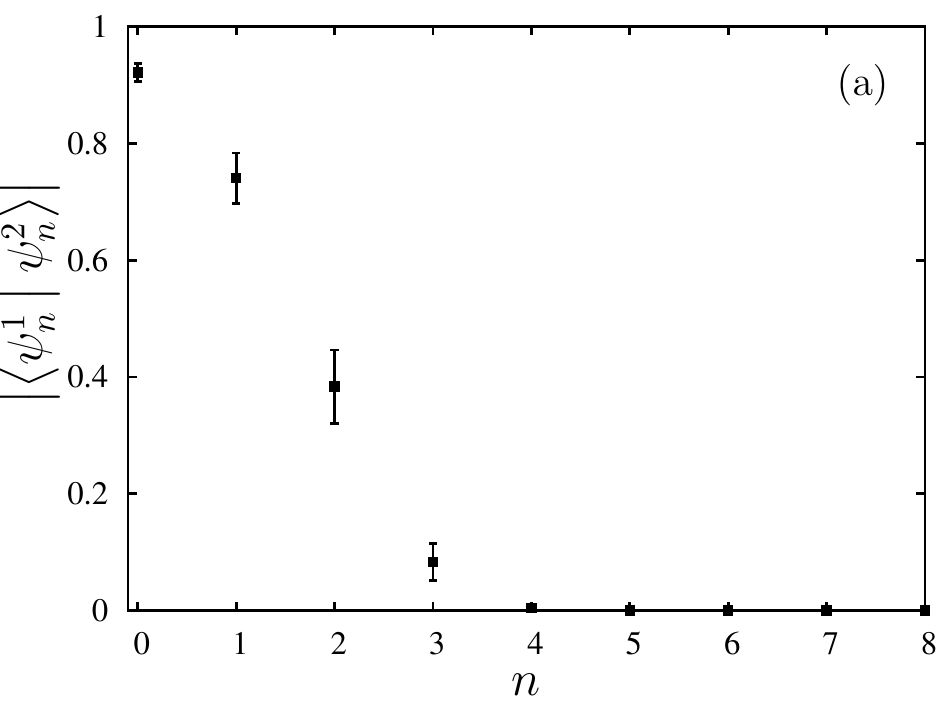}
\label{scalprod_a}}
\vspace{-2.0ex}
\subfigure{
\includegraphics[width=0.73\columnwidth]{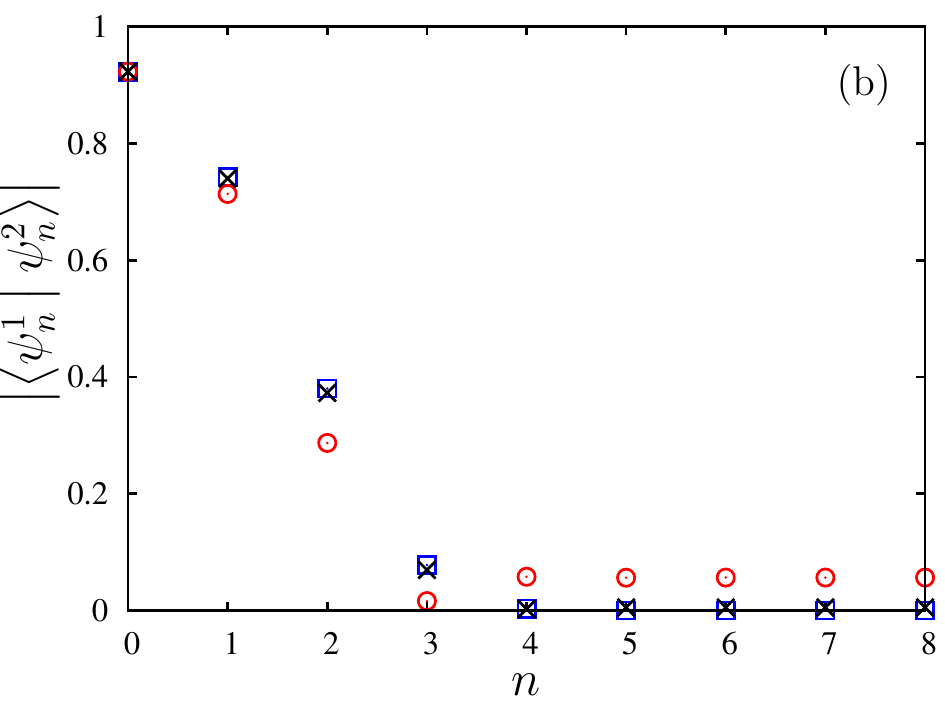}
\label{scalprod_b}} 
\caption{(color online) (a) The overlap of the states 
$\left|\psi^{1}_{0}\right>=0.98\left(\left|0\right>-0.2\left|1\right>\right)$ and 
$\left|\psi^{2}_{0}\right>=0.98\left(\left|0\right>+0.2\left|1\right>\right)$ after $n$ iterations of the ideal map, 
when there is an uncertainty described by a Gaussian distribution of standard deviation $\sigma=0.03$ around both the real and 
imaginary parts of the initial values $z_{1}=-0.2$ and $z_{2}=0.2$. The error bars represent the root-mean-square deviation from the mean 
(black squares) of the possible values of the scalar product. (b) The overlap of the states after $n$ iterations of the ideal map 
(blue squares), and the complete map with mean photon number 10 (red circles) and 100 (black crosses).}
\label{scalprod}
\end{figure}

The overlap of the above mentioned two initial states converges fast to zero, as we show in Fig.~\ref{scalprod}. To account for  possible imperfections in the preparation of the initial states we assumed a Gaussian uncertainty with a standard deviation of $\sigma=0.03$ in both the real and imaginary parts of the initial values  $z_{1}=-0.2$ and $z_{2}=0.2$. We note that this value of $\sigma$ assures that we sample from a  distribution of quantum states which have either positive or negative real part of the amplitude of state $\ket{1}$. Fig.~\ref{scalprod_a} shows that due to the nonlinear transformation the resulting uncertainty (represented by the error bars) in the initial value of the scalar product grows in the first and second step, but then decreases, and eventually becomes much smaller than its initial value (the error bars cannot be seen at the resolution of the figure for $n\geq 4$). Thus our procedure effectively discriminates between two different phases of small excitation amplitudes of the atoms. The evolution of the overlap of the above two initial states is not modified significantly when using the complete solution for the map, as can be seen in Fig.~\ref{scalprod_b}. Mean photon numbers of $\overline n = 10$ and $\overline n =100$ lead to essentially similar behavior to that of the idealized map (\ref{map}). Interestingly, the low-photon-number case leads to a faster decrease in the overlap during the first few steps of the iteration, but then converges to a larger value compared to the ideal map. 

\section{Experimental considerations}
\label{Section-experimental}

Our basic protocol involves atomic and photonic postselection and therefore there is always a
finite probability of failure. This means that in order to implement several iterations of the map,
one requires several copies of the initial qubit pair.  The procedure explained in 
Sec. \ref{onestep} has to be applied to every single copy of the ensemble. The number $N$ of qubit pairs 
required to achieve $n$ iterations, can be bounded from below by taking into account the
success probability $P_{\rm s}$ in \eqref{psuccess}. In addition, one has to take into account that half of the atoms
in the ensemble are lost after being measured. Therefore, the number of pairs scales exponentially as 
$N=(2/P_{\rm s})^n=(8/\cos^2\varphi)^n$.

On a first thought one would naively consider the use of $N$ optical cavities for $N$ atomic pairs. However,  there
is another simpler solution motivated by current experimental implementations \cite{Khudaverdyan,Brakhane}, where  
a standing-wave dipole trap or ``optical conveyor belt'' is used to coherently transport neutral atoms into an optical resonator. 

\begin{figure}
\includegraphics[width=0.47\textwidth]{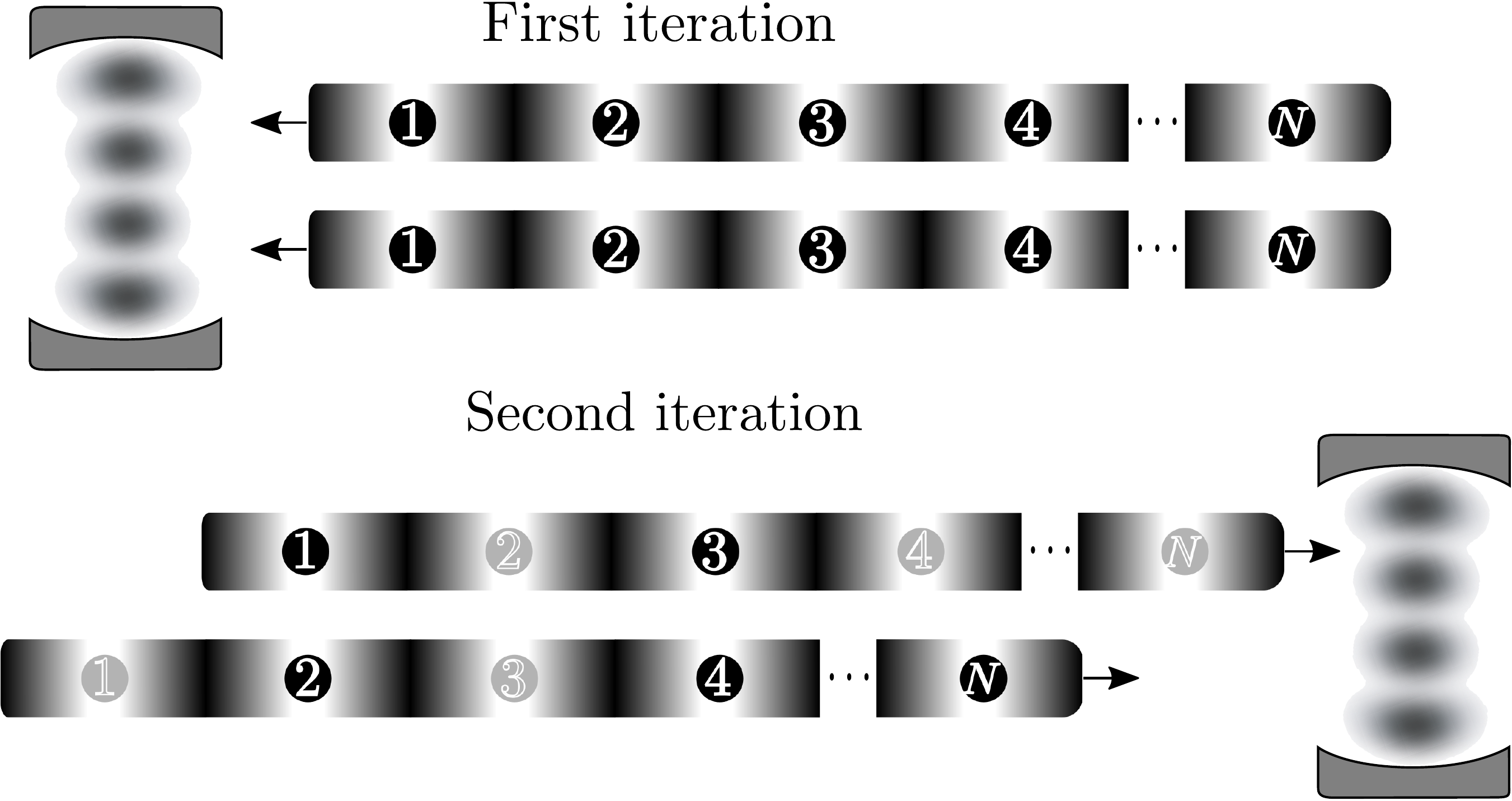}
\caption{\label{schemebelt}
Possible  implementation of the protocol using neutral atoms coherently transported using  optical conveyor belts.
}
\end{figure}

Using this setting, two conveyor belts are required to transport atoms into the cavity. In an initial stage, 
$N$ atoms are prepared in the minima of the
two optical traps and are aligned as depicted in Fig. \ref{schemebelt}. For convenience, we number the atoms
from left to right. The unitary gate $\hat U_\varphi$ is applied at this preparatory stage to atoms labeled with an
even (odd) number in the upper (lower) conveyor belt, we call them marked atoms.
The two conveyor belts are moved forward into the direction of
the cavity until the first pair reaches the other side of the cavity. Then, the conveyor belts stop in order to
allow the measurement of the first marked atom and the  field inside the cavity. Afterwards, 
the cavity is reset to the state $\ket\alpha$
and the conveyor belts move again repeating the process. After all atoms have interacted with the cavity, the marked atoms are 
blacklisted as they are no longer useful. 
They are depicted in gray in Fig. \ref{schemebelt}. In order to pair only the useful atoms, 
the lower conveyor belt is shifted one period
to the left, leaving the first marked atom without a partner. In this way, the potentially successfully prepared atoms
are aligned. The process is repeated with both conveyor belts moving to the opposite side to start the second iteration.
In the aforementioned implementation 
of the second iteration we have ignored the possibility of failure in the postselection. 
In order to overcome this problem, one has to keep track of successfully prepared atoms and then shift
the conveyor belts 
in order to align useful pairs before transporting them into the cavity.

Finally, it is worth noting that for the sake of simplicity we have only considered the dynamics
generated by the Hamiltonian $\hat H$ [see Eq. \eqref{Hamilton}] in the interaction picture with 
respect to the reference Hamiltonian $\hat H_0$ [see Eq. \eqref{H0}]. 
In the Schr\"odinger picture, or lab reference frame, the only differences are due to the free time evolution resulting from the Hamiltonian of Eq. (\ref{H0}). They lead to a relative phase
in the atomic state and they have to be taken into account in the field measurements (compare with Eq. \eqref{18}, for example). 
The one-atom state after a step of the protocol and up to normalization
can be written as $\ket{0}+f_\varphi(z)e^{i\phi-i\omega(t+t_1+t_2)}$\ket{1}.
Here we have considered a free evolution with time $t_1$ ($t_2$) before (after) the 
interaction which takes place for a time $t$. 
To prove this it suffices to note that $\hat H$ commutes with $\hat H_0$ and therefore
one can split the evolution operator in the Schr\"odinger picture  as
\begin{align}
  \hat{\mathcal U}&=
  e^{-i  \hat H_0 t_2/\hbar}
  e^{-i (\hat H+\hat H_0) t/\hbar}
  e^{-i \hat H_0 t_1/\hbar}\nonumber\\
  &=
  e^{-i \hat H_0(t+t_1+t_2)/\hbar}
  e^{-i \hat H t/\hbar}.
  \label{}
\end{align}
After the evolution, field and atom $B$ are projected into pure states
yielding for atom $A$ an evolution operator $\exp{[-i\omega\ketbra{1}{1}_A(t+t_1+t_2)]}$ which generates
the mentioned phase.
In order to keep the same form of the map in Eq. \eqref{map},  one could adjust the 
times in such a way that $t+t_1+t_2=2\pi/\omega$. 
Alternatively, one could eliminate this phase by driving the atoms with a classical electromagnetic field in a similar
way as we proposed to implement the gate $\hat U_\varphi$ in Eq. \eqref{gate}.

\section{Conclusion}

We have proposed a nonlinear map of qubit states in a cavity quantum
electrodynamical scenario where  the qubits are encoded in two-level atoms.  
The core step requires the interaction of two equally prepared atoms with the 
field inside an optical resonator according to the Tavis-Cummings model.  By
subsequent field detection and selective measurement of one of the atoms, the
unmeasured atom is postselected into a state nonlinearly depending on its 
initial state. 

From a mathematical point of view, we have studied the complex function
describing this mapping of pure qubit states, where we have exploited the
fact that any pure state of a qubit can be described by a complex parameter.
We have performed an analysis of stable cycles under the iteration of the
function and studied the behavior in the complex plane. In particular, we
have numerically investigated the Julia set which changes from connected to
disconnected for different parameters of the system. Thus, our study offers a
demonstration of chaotic behavior in a quantum mechanical setting involving
sequences of unitary transformations and postselective measurements. From a
physical perspective, we have proposed the realization of this scheme using
an ensemble of equally prepared atoms in two optical conveyor belts that are
coherently transported and interact in pairs with a single optical resonator.
We have estimated the number of atoms required for each iteration of the
protocol taking into account the success probability of the measurements
involved. Although possible realizations of this nonlinear qubit map require
cutting edge quantum technological developments, such as optical conveyor
belts and controlled two-qubit interactions with a single-mode radiation
field, in view of the rapid experimental advances in cavity quantum
electrodynamics its realization is within reach of nowadays technology.

The presented scheme provides an alternative approach to already established
quantum state discrimination protocols \cite{Paris}. We suggested an
effective implementation of the Schr\"odinger microscope in which two
initially close pure quantum states can be discriminated by amplifying the
distance between them and thus effectively orthogonalizing them. We have
shown that initial states of the two-level atoms with high overlap 
will become almost perfectly orthogonal by a few iterations of the scheme. Let 
us note that the orthogonalization procedure has a slightly different flavor 
than previous quantum state discrimination procedures. First, it is 
deterministic in the sense that there is a probability of success for the whole 
process, but then the resulting quantum state is fully determined by the initial
state. Second, it does not directly measure the orthogonalized systems, but rather prepares them in a non-demolition sense and therefore these systems can be used for further processing. Third, it is a purification
process as well, which naturally accounts for initial noise, and effectively
discriminates mixed nonorthogonal quantum states. Measurement-induced
nonlinear evolution in quantum mechanics is a concept which could be used by
other physical realizations of qubits to implement a Schr\"odinger
microscope.

\begin{acknowledgments}
This work was supported by the Hungarian Academy of Sciences (Lend\"ulet Program, LP2011-016) and the National Research, Development and Innovation Office (K115624, NN109651, PD120975) and by the Deutscher Akademischer Austauschdienst (M\"OB-DAAD project no. 65049). 
O. K. acknowledges support from the J\' anos Bolyai Research Scholarship of the Hungarian Academy of Sciences. 

\end{acknowledgments}

\end{document}